# 73.1: Large-Area Plasma-Panel Radiation Detectors for Nuclear Medicine Imaging to Homeland Security and the Super Large Hadron Collider

*Peter S. Friedman*
Integrated Sensors, LLC, Toledo, Ohio, USA

*Robert Ball, J. Wehrley Chapman, Daniel S. Levin, Curtis Weaverdyck, Bing Zhou*
Dept. of Physics, University of Michigan, Ann Arbor, Michigan, USA

*Yan Benhammou, Erez Etzion, M. Ben Moshe, Yiftah Silver*
Raymond and Beverly Sackler School of Physics and Astronomy, Tel Aviv University, Tel Aviv, ISRAEL

*James R. Beene and Robert L. Varner Jr.*
Oak Ridge National Laboratory, Holifield Radioactive Ion Beam Facility, Oak Ridge, Tennessee, USA

## Abstract

*A new radiation sensor derived from plasma panel display technology is introduced. It has the capability to detect ionizing and non-ionizing radiation over a wide energy range and the potential for use in many applications. The principle of operation is described and some early results presented.*

## 1. Introduction

A new class of highly pixelated, fast response, high gain, radiation detectors is being developed based on plasma panel technology. In its most basic form it is known as a plasma panel sensor or "PPS" [1]. By depositing a photocathode on an interior surface facing the gas, a light-sensitive PPS, known as a plasma panel photosensor or "PPPS" [2] can be realized with potential advantages over other high-gain, light-detection devices such as photomultiplier tubes (PMT), solid state photomultipliers (SSPM), gas electron multipliers (GEM), Geiger-mode avalanche photodiodes (APD), multichannel plate photomultipliers (MCP-PMT). By coupling the PPPS to a scintillator, a plasma panel scintillating detector ("PPSD") can be constructed for a host of applications: Compton telescopes [3], sampling calorimeters in high energy physics, medical imaging, homeland security, etc. The many potential attributes of PPS devices are attracting significant interest from nuclear physicists for detecting highly ionizing charged particles at radioactive ion beam (RIB) accelerators [3], as well as from high energy physicists for the detection of minimum ionizing particles (MIP) for the next generation of high and super-high luminosity colliders [4] such as the Super Large Hadron Collider (SLHC) at CERN and the International Linear Collider (ILC).

The goal of our research is to develop plasma panel based radiation detectors for both scientific and commercial applications. We describe below the basic theory of operation, our experimental effort and simulation results, and potential market opportunities for plasma display panel (PDP) manufacturers. For example, medical imaging, medical therapeutics and homeland security constitute multi-billion dollar markets that can provide order-of-magnitude higher margins than PDPs. In addition, manufacturing capitalization costs should be low as these detectors can ideally be fabricated in modified older generation PDP production facilities.

## 2. Plasma Panels as Radiation Detectors

A PDP comprises millions of cells per square meter, each of which, when provided with a signal pulse, can initiate and sustain a plasma discharge. Configured as a plasma panel detector, a PDP cell is biased to discharge when a free-electron is generated or ejected into the gas. The PPS as a re-engineered PDP, functions as a highly integrated array of parallel pixel-sensor-elements or cells, each independently capable of detecting free-electrons generated within the cell by incident ionizing radiation. Such electrons then undergo rapid electron avalanche multiplication that can be confined to the local pixel cell space. Because of the small electrode gap, large electric fields on the order of MV/m can arise with only a few hundred volts of bias. In effect the device thus constitutes a dense array of micro-Geiger plasma discharge cells. For example, for a minimum ionizing particle (MIP) detector, the source of free-electrons needed to initiate a discharge, results from the continuous energy loss (dE/dX) of the MIP (e.g. muon) as it traverses a drift region within the cell. However by appropriate design of the front-end cathode, a PPS can also detect neutrons and photons (e.g. photons ranging in energy from the visible to gamma rays) [1]-[4].

The potential attributes of a PPS include: fast response (~ 1 ns), high spatial resolution (< 100 µm), high gain (~ $10^6$), high data rate capability (> $10^9$ Hz/cm$^2$), low cost, low mass, radiation hardness and long lifetimes. Moreover these detectors can be hermetically sealed with no external gas system required. We anticipate that the PPS can be scaled to sizes commensurate with the needs of large detector systems, while being lightweight and structurally robust. One area of focus suggested by the above potential virtues of PPS detectors is in developing an advanced detector applicable to *relativistic* charged particle tracking in high energy physics – e.g. a MIP detector. Current MIP detectors have significant short-comings for their use in the next generation of colliders (e.g. SLHC) which the PPS could overcome. Beyond high energy physics, PPS devices should be well-suited for less demanding applications in nuclear physics such as detecting highly ionizing, *non-relativistic* (e.g. < 50 GeV) charged particles from radioactive ion beam (RIB) accelerators, or beam diagnostics in particle beam therapy (i.e. cancer treatment), etc.

Some potential features that distinguish the PPS technology are:

1) <u>Sparking and Gain</u>: A recurrent problem with micropattern detectors which operate with gains of ~ $10^3$-$10^4$ is destructive sparking. The PPS is designed to be a higher gain, Geiger-mode device and is intrinsically spark-free. An inline, current-limiting discharge resistor can be embedded within each cell to immediately drop the voltage at discharge, reducing the electric field to terminate the current pulse.

2) <u>Longevity</u>: PPS materials are glass, ceramics, non-reactive refractory/metal electrodes, and inert or non-corrosive gas mixtures. They contain no thin-film polymers used in other micropattern detectors, and no hydrocarbons that can degrade or outgas. The device lifetime should be orders-of-magnitude better than for Micromegas and gas electron multiplier (GEM) type detectors.

3) <u>Radiation Hardness</u>: PPS detectors employ radiation resistant materials and no internal semiconductors (readout electronics are externally located). They contain intrinsically radiation hard materials, namely inert gases, glasses, ceramics, metals.

4) <u>Spatial Resolution</u>: Full-color PDP's with a cell pitch of 100 µm were manufactured in the 1990's. A PPS with a similar cell pitch would have a spatial resolution of ~ 30 µm RMS.

5) <u>Stable and Uniform Response</u>: The sensitive electrostatics of the avalanche gap is determined by electrodes deposited on the same substrate with a high degree of uniformity. There is no critical dependence on the drift distance which is determined by the panel separation or by a mesh distance.

6) <u>High Rates</u>: Response is dependent on the formation of the PPS signal which is intrinsically fast with rise times expected to be of order 1 ns or less. Rate capability is limited by cell recovery times, expected to be on order of a few ns depending on specific cell geometry and discharge resistance and quenching of long-lived metastable species.

7) <u>Cost, Scalability and Electronics Readout</u>: Conceived as a re-engineered PDP, PPS detectors should benefit from well-established fabrication processes, materials, and associated mechanical and electronics infrastructure. The PPS readout electronics would be similar to those used in other high channel density, two coordinate detectors. The expected high gain of a PPS renders them intrinsically binary, possibly obviating an amplification stage and thus simplifying the front-end signal processing. Also because they are fabricated on glass substrates, high density, high speed, electrode-to-IC interconnections can be achieved via low cost, commercial technology such as that used for a variety of flat panel display and other high information content video/computer products.

## 3. PPS and PPPS Development Program

### 3.1. Technical Approach

A multifaceted experimental–theoretical approach is being pursued that is driven by the varied interests of the participating parties. Coordination of the overall effort is being provided by Integrated Sensors (I-S), with the university-based high energy physics MIP-PPS detector program centered at the University of Michigan (UoM), which has led the MIP-PPS modeling and simulation work, as well as the design effort for the readout electronics. The modeling effort is now transitioning from 2-D to 3-D simulations with much of the work shifting to Tel Aviv University (TAU). The PPS-RIB (radioactive ion beam) detector will be tested at Oak Ridge National Laboratory (ORNL), which will also evaluate the PPS efficacy for medical therapeutics (i.e. proton beam therapy) and homeland security (i.e. neutron detection). Modeling and testing of the PPPS for scintillator based applications, and for the PPS as a neutron detector is also centered at ORNL. The PPS and PPPS device work, including overall experimental/hardware design and fabrication, has been centered at I-S, which has also focused on the commercial applications.

To survey the general parameters of detector geometry, materials and gas mixture, a broadly configurable test chamber is under construction to allow the staging of a wide variety of PPS and PPPS test cell configurations and materials. The test chamber is a vacuum/pressure vessel with ports for gas supply and gas exhaust, cameras, spectrometers, and a variety of illumination, excitation and ion beam sources. The "open" cell design provides maximum experimental flexibility to map out virtually every PPS design parameter. The PPS/PPPS discharge gas is supplied via a gas mixing system that can provide up to a four component mixture with precise control of the lowest blended component down to ~ 0.01% for any fill pressure up to 5 atmospheres of positive pressure. A digitally-controlled motorized elevation stage, translatable along the vertical axis towards or away from the drift electrode or photocathode (i.e. fixed position front plate), serves as the platform upon which PPS/PPPS test cells (i.e. back substrates) are mounted – see Fig. 1a & 1b. The drift-field distance can be controlled over the range from ~ 50 µm to 1 cm with micron level precision. The first PPS/PPPS test devices use low-cost alumina back substrates, but will eventually use glass. To accurately measure rise times in the sub-nanosecond regime, an array of custom vacuum/pressure electrical feedthroughs has been designed to accommodate the hundreds of signal output lines, with each signal wire being a shielded 75Ω coaxial cable. Cable lengths are carefully controlled since a 1 mm variation in length corresponds to a signal propagation time difference of 3 ps.

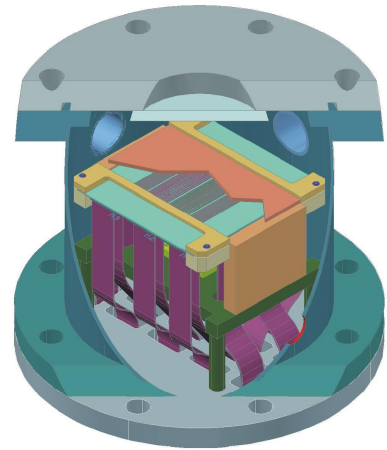

**Figure 1.a. (top view) PPS/PPPS Test Chamber with all four sectors of active cell area on back plate shown along with partial section of top (cover plate) drift electrode.**

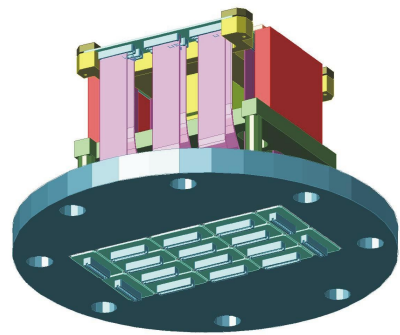

**Figure 1.b. (bottom view) PPS/PPPS Test Chamber with shielded coaxial cable and vacuum feedthroughs shown along with motorized elevation stage.**

The test chamber has been designed for a wide variety of entrance window materials, from ultra-thin (e.g. ~ 4 µm) foils to Cherenkov radiators, scintillators, VUV transparent crystals and glasses. For each gas mixture, approximately a half-dozen device operating pressures will initially be tested before evacuating and changing to another mixture. To maximize the device testing efficiency, the back substrate "top" surface has been divided into four different electrode resolution regions, covering a total active area of 53 cm$^2$ (see Figure 1a). Metal *vias* connect each top thin-film electrode to the bottom plane of the back substrate which uses thick-film conductors with controlled resistivity. Initially the cell discharge resistances are being printed on the bottom plane. The device front plate electrode is also interchangeable and can be an ultra-thin foil or any transparent or "metallized" electrode surface, photocathode, neutron converter, or other type of conversion layer.

In summary, the reconfigurable test chamber provides an extremely economical means to test almost any combination of PPS and PPPS device configurations and materials. A parallel effort, however, is also underway to fabricate low cost, discrete PPS and PPPS devices for specific applications. We think the latter effort, in particular, will benefit by the participation of one or more flat panel display manufacturers, especially those with PDP or field emissive display (FED) fabrication experience.

### 3.2. PPS – PPPS Cell Structure and Operation

All PDP products are designed so that the cell avalanche is self-limiting and self-contained. The total charge available to produce a signal is generally limited by the charge stored in the cell's internal capacitance. The stored charge thus effectively limits the maximum gain. The maximum possible gain therefore depends on details of cell geometry and materials and is estimated for a PPS with a 100 µm pixel pitch to be on the order of $10^6$. Since the cell is operated above the proportional mode it, in essence, may be considered to be a micro-Geiger counter. The signal pulse will be independent of the number of initiating free electrons, rendering therefore the PPS as intrinsically digital. The gain may be sufficient to simplify or even obviate signal amplification electronics.

Efficient MIP detection requires an electrode layout with a vertical drift region of ~ 2 to 3 mm coupled to a suitably matched transverse electric field avalanche region. The drift region is required to ensure sufficient probability that a passing particle will produce at least one ion-pair. For example, in Ar at 1 atmosphere, a MIP produces ~ 30 interaction "clusters" per cm yielding at least one ion-pair [5]. The cluster generation is Poisson distributed, thus in a 2 mm drift region at 500 torr, the probability to generate at least a single cluster with, on average 3 ion-pairs, is about 99%. Due to the large gain, a single ion-pair can initiate a signal, suggesting that a drift region of 2 to 3 mm is sufficient to produce signals with high efficiency.

The PPS is a DC device, so it most closely resembles a DC-PDP. Important advantages to the DC structure include low pixel capacitance for fast timing and low power consumption, and the near elimination of surface wall charge for maximum field bias. A *conceptual* representation of the cell structure is shown in Figure 2. In this depiction the cell is represented by the proximity of the two electrodes (X and Y) defining the discharge region. The Z-strip electrodes provide orthogonal coordinate readout running transverse to the discharge cells and capacitively coupling to them. The high voltage on the discharge (i.e. "X") electrode is supplied via a resistive element connecting it to a high voltage bus. The resistance drops the high voltage at discharge, effectively terminating it. The local cell resistance also provides electrical isolation by preventing adjacent channels from dumping their stored charge through the discharging cell. In this sense the operation of the PPS resembles a resistive plate chamber (RPC) type detector. The effectiveness of this resistance in isolating the discharge is investigated with SPICE simulations as described in Section 3.3 below.

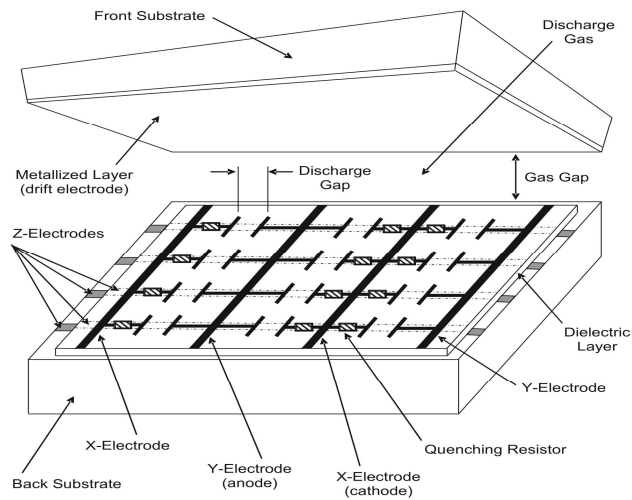

**Figure 2. Representation of 4-electrode PPS structure. The dielectric layer can be eliminated by using a thin (~ 1 mm) back substrate (drawing is conceptual only & not to scale).**

### 3.3. Modeling and Simulations

The feasibility of the PPS has been demonstrated to the U.S. Dept. of Energy (DOE) and the U.S. Defense Threat Reduction Agency (DTRA) using modified AC-PDP's and DC-PDP's, which were able to detect the gamma-ray emitting radioisotopes: $^{57}$Co, $^{99m}$Tc and $^{137}$Cs [1]. In addition to the experimental efforts currently underway, device development has been guided by modeling and simulations. Gas ionization, drift electron generation, electron transport and arrival time distributions are computed using Garfield and associated Magboltz and HEED programs. Garfield can also be used to model the electrode geometry and resultant electric field, although greater descriptive flexibility is obtained from Maxwell-2D. These simulations have allowed us to analyze both the electron and ion drift field shapes, electron path/time dispersion and the rise time jitter for about a dozen PPS cell configurations, including various discharge gases, cell pitches, drift electrode distances, discharge and sense electrode widths and discharge gaps, and different heights of the discharge electrode extending into the drift field. It is clear from these simulations that the options for electrode design promise a wide range of field shaping choices. A newer addition to the modeling software suite is the COMSOL Multiphysics package. This package enables 3-D electrostatic modeling which is essential to accurately describe the PPS architectures. Simulations have also been performed for various PPPS configured Compton telescope arrangements using GEANT3 and GEANT4 from CERN [3].

Output from the simulation packages provides detailed description of the cells' electric field, energy density matrix and electrode pair capacitances. The capacitances thus obtained include those of the discharge electrode and all stray or parasitic capacitances presented by the complex electrode array. A detailed circuit analysis, which yields information on signal development, cross-talk and power consumption, is performed using the SPICE circuit simulator. SPICE modeling is predicated on the electric field and capacitance results obtained from Maxwell, Garfield, etc.

The electrode configurations have been modeled using Garfield and the Maxwell-2D programs. These calculate the electric fields, equipotential surfaces and electrode capacitances. The sense electrode (Y) capacitance was determined to be ~ 2.5 fF, assuming a glass dielectric of permittivity 5.5. The SPICE modeling and analysis package was then invoked to compute the temporal profile of the high voltage X-discharge lines (Figure 3) and the sense signals (Figure 4) for a 100 μm cell pitch with 3 mm drift-field distance. A circuit model was employed based on a chain of 13 cells, including the capacitances of the cell electrodes, orthogonal Z-strip electrodes and various stray capacitances, and all discharge line resistances. The signal termination resistance was 120 Ω and the intrinsic cell capacitance was estimated at 3 fF.

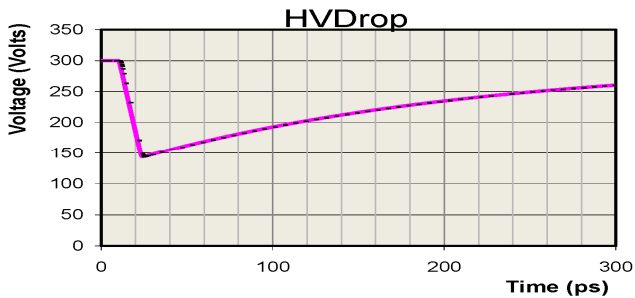

**Figure 3. Time profile of high voltage drop across the hit cell in a 13 multi-cell chain. The rise and fall times reflect the cell capacitances and resistances. The drop to ½ the cell potential occurs with a rise time of ~ 8 ps. The 1/e fall time is ~ 250 ps.**

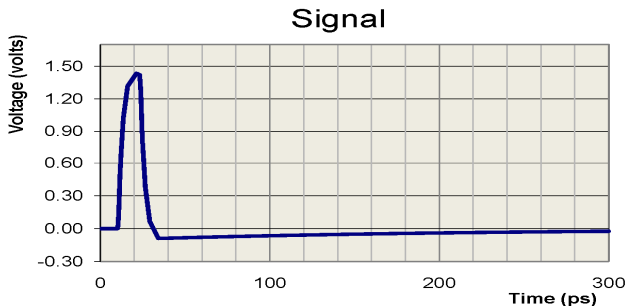

**Figure 4. Time profile of sense line signal produced by hit cell in a 13 multi-cell chain, across a 120 Ω output impedance.**

The SPICE simulation has the following assumptions. The voltage needed to extinguish the discharge occurs at ½ the bias voltage. In accordance, the charge released by the current source was set to a magnitude corresponding to ½ the stored charge on the cell. The duration of the current pulse is determined by avalanche formation across the small discharge gap and assumed to be a near delta function. A number of conclusions follow:

1) The potential across an activated pixel drops to ½ the supply value in ~ 10 ps, assuming a delta-function current source.
2) No other cell in the chain, including no neighboring cells, experience any drop of the high voltage. Therefore no stored charge from neighbor cells contributes to the hit cell signal, which thus remains electrically isolated during the discharge.
3) The sense signal has a magnitude of 1 volt. The Z-strip signal, while smaller than the primary sense line signal, should still be hundreds of mV, but of opposite polarity.
4) The recovery time constant is ~ 250 ps. Thus the duration for full cell recovery is of order ns. The recovery time is directly but not exclusively dependent on the inline pull-up resistance.
5) The integrated energy dissipation for a single pixel discharge is ~ 56 pJ, which sets the power consumption, excluding the readout electronics, at 2 μW/cm$^2$.
6) The fast response of the PPS suggests use as trigger elements when adjacent layers are combined to define a trajectory compatible with particles of interest.

## 4.     Impact and Conclusions

PPS/PPPS detectors offer potentially higher signal rates, faster timing, and more precise positional information than other ionization sensing devices but in many ways are similar, particularly with respect to triggering and readout possibilities. Large signals also distinguish PPS/PPPS devices from those that require high gain amplifiers prior to the hit processing. Radiation hardness, scalability and low cost are important attributes. Initial feasibility experiments, simulations and analysis for these devices are very encouraging.

The potential impact of the PPS and PPPS includes an extremely broad range of commercial and scientific applications, such as: time-of-flight (TOF) PET/CT, gamma cameras, multislice-CT imagers, SPECT/MRI detectors, X-ray imagers, detection of nuclear materials via neutron emission and high explosives by active interrogation, large MIP detectors, TOF Cherenkov detectors, focal plane tracking detectors, neutron calorimeters, active pixel beam monitors, Compton telescopes, non-destructive materials/structural testing and inspection, etc. Given the breath of applications and possible order-of-magnitude benefits, the commercial impact and potential benefits of the PPS and PPPS technology could be both profound and transformational.

## 5.     Acknowledgements

This work has been supported in part by both Phase-I and Phase-II SBIR grants from the U.S. Department of Energy, and by a grant from the United States – Israel Binational Science Foundation.

## 6.     References

[1] Peter S. Friedman, "A New Class of Low Cost, High Performance Radiation Detectors", IEEE Nuclear Science Symposium Conference Record (Puerto Rico), 2815-2822 (2005).

[2] Peter S. Friedman, "Plasma Panel Sensors as Scintillation Detectors", IEEE Nuclear Science Symposium Conference Record (San Diego), 1150-1159 (2006).

[3] Robert L. Varner and James R. Beene, "Simulation of a Scintillator-based Compton Telescope with Micropattern Readout", IEEE Nuclear Science Symposium Conference Record (Honolulu), 1248-1254 (2007).

[4] R. Ball, J. W. Chapman, E. Etzion, P. S. Friedman, D. S. Levin, M. Ben Moshe, C. Weaverdyck and B. Zhou, "Plasma Panel Detectors for MIP Detection for the SLHC and a Test Chamber Design", IEEE Nuclear Science Symposium Conference Record (Orlando), 1321-1327 (2009).

[5] C. Amsler *et al*. The Review of Particle Physics, 28. Particle Detectors, Physics Letters, B667, **1**, p.28 (2008). Also F. Sauli, "Principles of Operation of Multiwire Proportional and Drift Chambers", CERN 77-09, Geneva (1977).